\documentclass[aps,prd,reprint,superscriptaddress]{revtex4-1}

\usepackage{graphicx}
\usepackage{amssymb, amsmath,mathtools,amsthm}
\newtheoremstyle{dotless}{}{}{\itshape}{}{\bfseries}{}{ }{}
\theoremstyle{dotless}
\makeatletter
\def\@endtheorem{\endtrivlist}
\makeatother
\newtheorem*{proposition*}{}

\usepackage[colorlinks=false,urlcolor=black]{hyperref}

\newcommand{\be}{\begin{equation}}
\newcommand{\bea}{\begin{align}}
\newcommand{\eea}{\end{align}}
\newcommand{\beq}{\begin{equation}}
\newcommand{\ee}{\end{equation}}
\newcommand{\eeq}{\end{equation}}

\newcommand{\vfr}{v^{\hat{r}}_F}
\newcommand{\vfh}{v^{\hat{\theta}}_F}
\newcommand{\vfp}{v^{\hat{\phi}}_F}

\newcommand{\apjl}{{ApJ Lett.}}

\newcommand{\mnras}{{MNRAS}}

\newcommand{\ssr}{{Space~Sci.~Rev.}}

\newcommand{\jgr}{{J.~Geophys.~Res.}}

\def\ip{${\cal I}^+$}
\def\im{${\cal I}^-$}

\begin{document}


\title{Energy extraction from boosted black holes: 
Penrose process, jets,\\ and the membrane at infinity}



\author{Robert F. Penna}
\email[]{rpenna@mit.edu}
\affiliation{Department of Physics and Kavli Institute for Astrophysics and Space Research,
Massachusetts Institute of Technology, Cambridge, Massachusetts 02139, USA}


\date{\today}

\begin{abstract}

Numerical simulations indicate that black holes carrying linear momentum and/or orbital momentum can power jets.  The jets extract the kinetic energy stored in the black hole's motion.  This could provide an important electromagnetic counterpart to gravitational wave searches.  We develop the theory underlying these jets.  In particular, we derive the analogues of the Penrose process and the Blandford-Znajek jet power prediction for boosted black holes.  The jet power we find is $(v/2M)^2 \Phi^2/(4\pi)$, where $v$ is the hole's velocity, $M$ is its mass, and $\Phi$ is the magnetic flux.  We show that energy extraction from boosted black holes is conceptually similar to energy extraction from spinning black holes.  However, we highlight two key technical differences: in the boosted case, jet power is no longer defined with respect to a Killing vector, and the relevant notion of black hole mass is observer dependent.    We derive a new version of the membrane paradigm in which the membrane lives at infinity rather than the horizon and we show that this is useful for interpreting jets from boosted black holes.  Our jet power prediction and the assumptions behind it can be tested with future numerical simulations.

\end{abstract}

\pacs{}

\maketitle

\section{Introduction}
\label{sec:intro}

Recent numerical simulations \cite{
2009PhRvL.103h1101P,
2010PhRvD..81h4007P,
2010Sci...329..927P,
Palenzuela:2010xn,
2011PNAS..10812641N,
2012ApJ...749L..32M,
2012ApJ...754...36A,
2013PhRvD..88b1504P} 
and analytic estimates \cite{2011PhRvD..83f4001L,2011ApJ...742...90M,2013PhRvD..88f4059D,2014PhRvD..89j4030M} suggest black holes carrying linear and orbital momentum can power jets .
The jets are driven by electromagnetic fields tapping the kinetic energy stored in the black hole's motion.  The power of the simulated jets scales approximately as $v^2$, where $v$ is the hole's velocity \cite{2011PNAS..10812641N}.   Such jets could be an important electromagnetic counterpart to gravitational wave signals because $v\sim 1$ in the final stages of black hole-neutron star and black hole-black hole mergers.  This paper develops the theory underlying these jets.

Our first goal is to develop the analogue of the Penrose process \cite{1969NCimR...1..252P,1992mtbh.book.....C} for boosted black holes.  The original Penrose process is a simple mechanism for extracting rotational energy from Kerr black holes.  It relies on the fact that certain geodesics near spinning black holes have negative energy (with respect to global time).   In the original Penrose process, a particle with positive energy travels toward the black hole and decays into two daughter particles.  One of the daughter particles falls into the black hole with negative energy and the other returns to infinity.  The final particle has more energy than the original and the black hole's mass decreases.  

We derive the analogous process for boosted Schwarzschild black holes in Sec. \ref{sec:penrose}. In the rest frame of a Schwarzschild black hole there are no negative energy trajectories and it is impossible to lower the black hole's mass via the Penrose process.  However, in a boosted frame (where the black hole carries linear momentum), there are negative energy trajectories.   We use these trajectories to derive the analogue of the Penrose process. This gives a simple example of energy extraction from boosted black holes.  It may be useful for describing the interactions of stars with moving black holes.

Our second goal is to develop the analogue of the Blandford-Znajek (BZ) model \cite{1977MNRAS.179..433B,1986bhmp.book.....T}.  In the original BZ model, electromagnetic fields tap a spinning black hole's rotational energy and drive jets.  
The BZ jet power prediction is currently being tested against astrophysical observations of spinning black holes \cite{2013SSRv..tmp...73M,2013ApJ...762..104S}.  

We develop the analogue of the BZ jet power prediction for boosted black holes in Sec. \ref{sec:jets}.  For small $v$, we find
\beq\label{eq:main}
P_{\rm jet}  = \frac{1}{4\pi}\left(\frac{v}{2M}\right)^2\Phi^2,
\eeq
where $\Phi$ is the magnetic flux at infinity and $M$ is the black hole's rest mass.  This is similar to the BZ prediction for spinning black holes but with $v/(2M)$ in place of the horizon angular velocity $\Omega_H$, the flux evaluated at infinity rather than the horizon, and a slightly different normalization constant.  The $v^2$ scaling  is consistent with earlier simulations \cite{2010Sci...329..927P,Palenzuela:2010xn,2011PNAS..10812641N,2013PhRvD..88b1504P} and estimates \cite{2011PhRvD..83f4001L,2011ApJ...742...90M,2013PhRvD..88f4059D}. Our formula predicts jets from boosted black holes and spinning black holes have comparable strength when $v/(2M)\sim \Omega_H$.   Numerical simulations suggest the true power of jets from boosted black holes is lower by as much as a factor of 100 \cite{2011PNAS..10812641N}.  We discuss possible reasons for this discrepancy  in Sec. \ref{sec:jets} but save a detailed comparison  for the future.

Our third goal is to develop a new version of the membrane paradigm in which the membrane lives at future null infinity, \ip.  In the usual membrane paradigm, the membrane lives at the black hole horizon \cite{1986bhmp.book.....T,Parikh:1997ma} and energy extraction is driven by torques acting on the membrane \cite{1986bhmp.book.....T,2013MNRAS.436.3741P}.  However, the energy flux at the horizon of a boosted black hole is not expected to match the energy flux at \ip\ in our jet model.  So it is more natural to place the membrane at infinity.  We derive this new version of the membrane paradigm in Sec. \ref{sec:membrane}.   Energy extraction from boosted black holes may be formulated in terms of interactions with the membrane at infinity.  Ordinary BZ jets and other processes involving black holes may also be reinterpreted using this formalism.

The idea of reformulating black hole physics in terms of a fluid at infinity (or perhaps a ``screen''  some finite distance outside the horizon) is not new.  The idea has been developed extensively for asymptotically anti--de Sitter black holes \cite{Myers:1999psa,2008IJMPD..17.2571H,Emparan:2009at,2011arXiv1107.5780H} and it has also been applied to asymptotically flat black holes \cite{Freidel:2014qya}.    The main novelties of our approach are to emphasize the connection with the classical black hole membrane paradigm and to develop the electromagnetic properties of the membrane at infinity which are important for describing jets.  

To summarize, in Sec. \ref{sec:penrose} we derive the analogue of the Penrose process for boosted black holes,  in Sec. \ref{sec:jets} we derive the analogue of the BZ model, and in Sec. \ref{sec:membrane} we derive a new version of the membrane paradigm in which the membrane lives at infinity.  We use this formalism to give an alternate interpretation of jets from boosted black holes.  We summarize our results and discuss open problems in Sec. \ref{sec:conc}.  Supporting calculations are collected in Appendices \ref{sec:boostinv}-\ref{sec:outgoing}.

\section{Boosted Black Holes and Penrose Process}
\label{sec:penrose}

\subsection{ADM 4-momentum}
\label{sec:adm}

The Schwarzschild metric in Kerr-Schild (KS) coordinates, $(\tau,x,y,z)$, is
\beq\label{eq:gKS}
g_{\mu\nu} = \eta_{\mu\nu} + 2 H l_\mu l_\nu,
\eeq
where $H=M/r$, $r=\sqrt{x^2+y^2+z^2}$, and $l_\mu  = (1,x/r,y/r,z/r)$.
To obtain the boosted solution, set \cite{2002PhRvD..66h4024H,2010nure.book.....B}
\begin{align}
d\tau &= \gamma (d\tau' - v dz'),\label{eq:tp}\\
dz &= \gamma (dz' - v d\tau'), \label{eq:zp}\\
dx &= dx',\\
dy &= dy',\label{eq:yp}
\end{align}
where $v$ is a constant parameter, $0<v<1$, and $\gamma = 1/\sqrt{1-v^2}$.
In the boosted frame, $(d\tau',dx',dy',dz')$, the black hole is moving in the $+z$ direction.  
The boosted metric is a solution of the vacuum Einstein equations because it is related to the Schwarzschild solution by a coordinate transformation \eqref{eq:tp}-\eqref{eq:yp}.  The horizon is at $r=2M$.  Its area is invariant under the boost but its shape is distorted: it becomes squashed along the direction of motion \cite{Huq:2000qx}.

If spacetime is foliated with respect to $\tau$, then the black hole's ADM 4-momentum is
\beq
P^{\rm ADM}_\mu = (-M,0,0,0).
\eeq
If spacetime is foliated with respect to $\tau'$, then its ADM 4-momentum is 
\beq
P^{\rm ADM}_{\mu'} = (-\gamma M,0,0,\gamma M v).
\eeq
That is, the black hole has linear momentum $\gamma M v$ in the boosted frame.  These are standard calculations, see for example \cite{2010nure.book.....B}. 

The black hole's energy in the boosted frame, $\gamma M$, is larger than its energy in the unboosted frame by a factor of $\gamma$.  However, in both frames the black hole's irreducible mass is $M_{\rm irr}=(\mathcal{A}/16\pi)^{1/2}=M$.  This follows from boost invariance of the horizon area, $\mathcal{A}$ (see Appendix \ref{sec:boostinv} for a proof).  So in the black hole's rest frame its ADM energy and irreducible mass coincide, but in the boosted frame they do not.  The difference,
\beq
\gamma M-M_{\rm irr} = (\gamma-1) M,
\eeq
is the energy that can be extracted from the boosted black hole.  

Energy extraction from  a boosted black hole is an observer-dependent process because -$P^{\rm ADM}_{\tau'}$ is not a Lorentz invariant.  What one observer interprets as energy transfer from black hole to matter, another observer interprets as energy transfer from matter to black hole.  However, the boosted picture is more natural for astrophysical problems involving kicked and orbiting black holes.   It is also conceptually interesting.    Rotational energy extraction from Kerr black holes is an observer independent process because $-P_{\rm ADM}^2$, a Lorentz invariant, decreases.

\subsection{Ergosphere}

\begin{figure}
\hspace*{-0.3in}
\includegraphics[width=0.95\columnwidth]{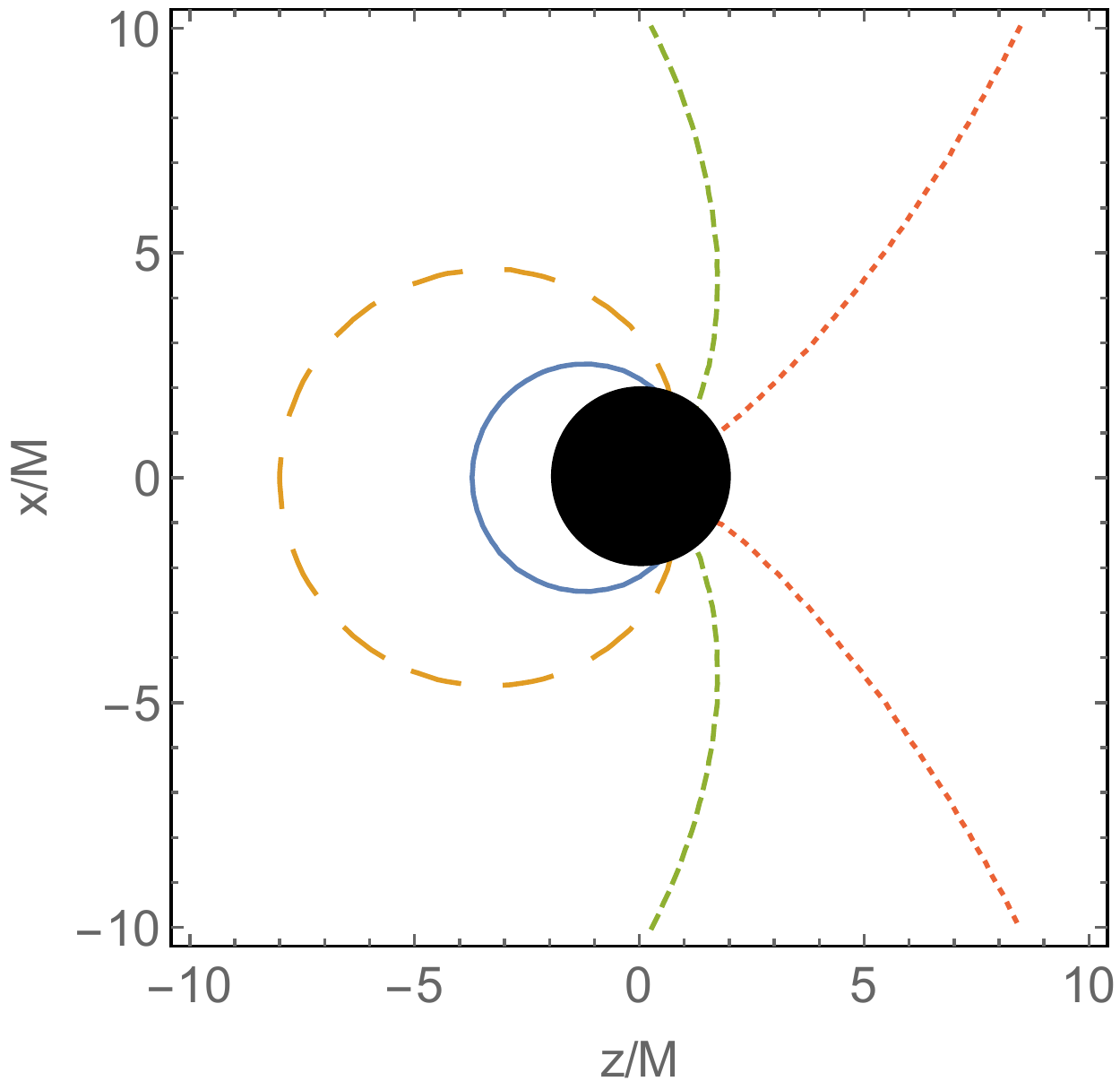}
\caption{Ergosphere for a boosted Schwarzschild black hole moving in the $+z$ direction with velocity $v=0.3$ (solid blue), $0.6$ (long dashed orange), $0.9$ (dashed green), and $0.99$ (dotted red).  The event horizon of a $v=0$ black hole is shown for comparison.  The ergosphere lags behind the event horizon.}
\label{fig:ergosphere}
\end{figure}

The ergosphere of a boosted Schwarzschild black hole is a coordinate dependent concept because $\partial_{\tau'}$ is not Killing.  Nonetheless, defining the ergosphere in a natural coordinate system gives insight into general features of energy extraction from boosted black holes.  In boosted KS coordinates, the ergosphere is the region where $\partial_{\tau'}$ is spacelike, or 
\beq
g_{\tau'\tau'} = \gamma^2 (g_{\tau\tau}+v^2 g_{zz}-2v g_{zt})>0.
\eeq  
Plugging in \eqref{eq:gKS} gives the radius of the ergosphere,
\beq
r_{\rm static} = 2M\gamma^2(1-v\cos\theta)^2.
\eeq
Observers inside the ergosphere cannot remain at rest with respect to $\partial_{\tau'}$.   Fig. \ref{fig:ergosphere} shows the ergosphere for several values of $v>0$.  The ergosphere is offset from the black hole and extends to
\beq
r_{\rm static}(0) = 2M \frac{1+v}{1-v}. 
\eeq
For $v=0$, the ergosphere coincides with the horizon.  For $v\rightarrow 1$, it extends to infinity. 
This is in marked contrast with the situation for Kerr black holes,  for which the ergosphere is always centered on the horizon and confined within $r\leq 2 M$.

\subsection{Penrose process}

A classic example of rotational energy extraction from Kerr black holes is the Penrose process  \cite{1969NCimR...1..252P,1992mtbh.book.....C}.  In this process, a particle falls into the ergosphere of a Kerr black hole and splits in two.  One of the daughter particles falls into the black hole along a negative energy geodesic and the other returns to infinity.  The final particle has more energy than the original particle and the black hole loses mass.
It is useful to work out the analogous process for  boosted black holes.  This is a warm-up for the more challenging problem of understanding black hole jets.  It may also be relevant for describing the interactions of stars with moving black holes.

Consider a particle with 4-momentum 
\beq
u_{\mu'} = (u_{\tau'},u_{x'},u_{y'},u_{z'})
\eeq
and energy $E'=-u_{\tau'}$ in the boosted frame \eqref{eq:tp}-\eqref{eq:yp}.  In this frame the black hole carries momentum along $z$.  A coordinate transformation gives
\beq
E' = \gamma (E + v u_z),\label{eq:eboost}
\eeq
where $E=-u_\tau>0$ and $u_z$ are the particle's energy and momentum in the black hole rest frame.  The boosted energy $E'$ is negative when $v u_z<0$  and $|vu_z|> E$.  The first condition means the particle and the black hole travel in opposite directions along $z$.  If such a particle is accreted, then the black hole's energy increases by $\gamma E$ (because it adds the particle's unboosted frame energy to its own), and it decreases by $-\gamma v u_z$ (because it loses kinetic energy).  The condition $|vu_z|> E$ means the latter effect wins.  This is impossible in flat spacetime, where
\beq\label{eq:onshell}
E = \sqrt{m^2 + u_x^2 + u_y^2 + u_z^2} \geq |u_z|.
\eeq
However, in black hole spacetimes, \eqref{eq:onshell} is replaced with $g^{\mu\nu}u_\mu u_\nu=-m^2$, and $|vu_z|> E$ is possible.  Roughly speaking, the gravitational field can contribute a negative potential energy to $E$. 
In the boosted Schwarzschild metric, particles at infinity have $E'\geq 0$ but particles at finite radii may have $E'<0$.

So we are led to consider something like the original Penrose process.  A positive energy ($E'>0$) particle at infinity falls toward a boosted black hole and splits in two.  One half follows a negative energy trajectory into the hole and the other escapes to infinity. The negative energy particle must move against the direction of the black hole's motion and be gravitationally bound.   The outgoing particle will have more energy than the original and the black hole will lose energy.

We have found numerical solutions for this process.  Assume the particles move in the $xz$-plane, so $u_{y}=0$.  Each trajectory is then fully characterized by three constants: $u_\tau$, $u_z$, and rest mass $m^2=-u^\mu u_\mu$.  The trajectories cannot be geodesics because $\partial_{z}$ is not Killing, but they could be achieved by using rocket engines to adjust a freely falling particle's momentum along $x$.

Fig. 1 shows one of our solutions.  
Particle $A$, with $u_\tau=-3/2$, $u_z = 0.9867$, and $m^2=-1$, travels from infinity to the interaction point $(r_*,\theta_*)=(4M,\pi/8)$. There it splits into massless daughter particles $B$ and $C$.  
Particle $C$ falls into the black hole with $u_\tau=-1/5+\epsilon$ and $u_z=-1/5$, where $\epsilon=10^{-4}$.  
Particle $B$ returns to infinity with 4-momentum fixed by momentum conservation at the interaction point:
$u^B_\mu=u^A_\mu-u^C_\mu$ at ($r_*,\theta_*$).

\begin{figure}
\includegraphics[width=\columnwidth]{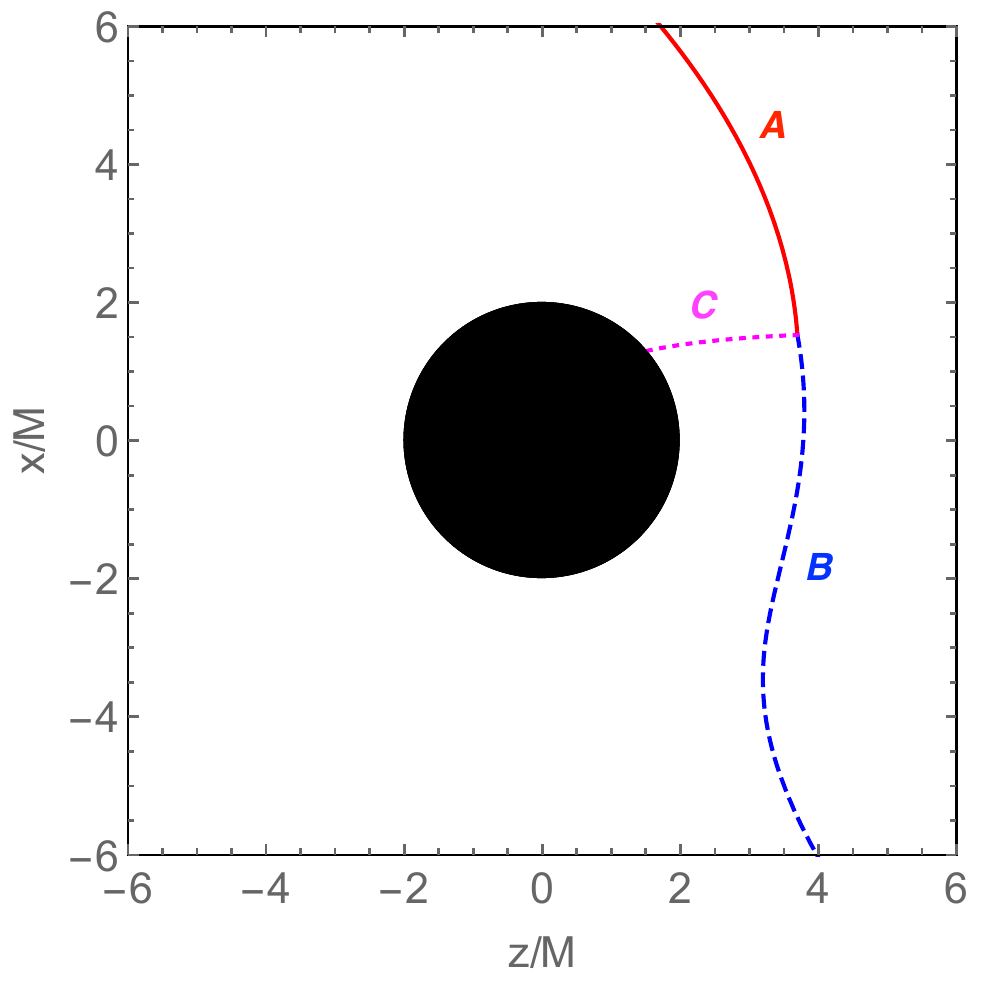}
\caption{Penrose process for a boosted Schwarzschild black hole moving in the $+z$ direction with velocity $v$.  Timelike particle $A$ (solid red) travels from infinity toward the black hole.  At the interaction point $(r_*,\theta_*)=(4M,\pi/8)$ it splits into massless daughter particles $B$ (dashed blue) and $C$ (dotted magenta).  $B$ returns to infinity and $C$ falls into the black hole.  For $v$ near $1$, $C$ has negative energy and $B$ returns to infinity with more energy than $A$. }
\label{fig:penrose}
\end{figure}

The boosted frame energies \eqref{eq:eboost} are
\begin{align}
E'_A &= \gamma (3/2 + 0.9867 v),\label{eq:EA}\\
E'_B &= E'_A - E'_C,\label{eq:EB}\\
E'_C &= \gamma (1 - v - 5\epsilon)/5.\label{eq:EC}
\end{align}
If  $v$ is near $1$, then $C$ falls into the black hole with  $E'_C<0$ and $B$ returns to infinity with $E'_B>E'_A$.   This is a concrete example of energy extraction from boosted black holes.    Further details of our method for finding these solutions and a second example are given in Appendix \ref{sec:penroseapp}.

There may be situations where this process is astrophysically relevant.  One can imagine a binary star $A$ that splits apart and creates a hypervelocity star $B$.  Or $A$ could be a single star that is tidally disrupted into streams $B$ and $C$.  We leave further discussion of these problems for the future.

One difference between our solutions and the usual Penrose process is that we consider nongeodesic trajectories, while the usual Penrose process describes geodesics.  External forces are required to keep particles on our trajectories.  This could make it difficult to distinguish whether the energy extracted to infinity is derived from the black hole or from the external forces.  However, we do not believe this is a problem.  The energy extracted to infinity in our solutions exactly matches the negative energy carried into the black hole.  So the black hole's energy decreases by the same amount as the energy gained, and it is fairly clear that the black hole is the source of energy.

The use of nongeodesic trajectories was forced upon us by the fact that linear momentum is not conserved along geodesics in the Schwarzschild metric.  A simple thought experiment illustrates the difficulty. Suppose a particle is dropped from rest into a nonmoving Schwarzschild black hole along a geodesic.   The initial momentum of the system is zero.  The particle speeds up as it falls toward the hole and crosses the horizon with nonzero linear momentum.  So the final momentum of the black hole appears to be nonzero, violating momentum conservation.  
One way to avoid this problem would be to do a fully general relativistic calculation incorporating the fact that as the particle falls toward the hole, the hole also falls toward the particle.  This goes beyond the scope of this paper.   
An alternate approach, which we chose, is to use nongeodesic trajectories that conserve linear momentum.  This seems to give the closest analogue to the usual Penrose process for test particles interacting with boosted astrophysical black holes.

Boosted Schwarzschild black holes are related to static Schwarzschild black holes by a Lorentz boost.  The Penrose process for static Schwarzschild black holes is impossible, so it may seem puzzling that the Penrose process exists for boosted black holes.  A helpful analogy is the billiards problem of scattering a cue
ball off of an eight ball.  In one frame, the eight ball is at rest
and gains energy from the cue ball, while in another frame the cue
ball is at rest and gains energy from the eight ball.  
Both descriptions are physically equivalent, the point being that the
energy, defined as the time component of four-momentum, is not a
Lorentz invariant.

Similarly, in the black hole rest frame the particles lose energy to the black
hole and there is no Penrose process.  However, in the boosted frame the black hole's energy (defined as the
time component of its ADM 4-momentum) is larger than its irreducible
mass, and it can transfer energy to the particles.

\subsection{Boosted black strings}

The metric 
\beq
ds^2 = -\left(1-\frac{r_+}{r}\right) dt^2 + \frac{dr^2}{1-r_+/r} + r^2 d\Omega_2^2 + dz^2.
\eeq
is a black string in 4+1 dimensions \cite{Horowitz:1991cd}.  It is a solution of the 5d vacuum Einstein equations.  The horizon is at $r=r_+$ and has topology $S^2\times \mathbb{R}$.  

A boosted black string may be obtained using the Lorentz transformation \eqref{eq:tp}-\eqref{eq:zp}.  Now the string carries momentum along $z$.  The ergosurface is at \cite{Hovdebo:2006jy}
\beq
r_{\rm static} = \gamma^2 r_+.
\eeq
Since $\partial_z$ is Killing, one might expect to find Penrose process solutions using particles following geodesics.  This would make  the Penrose process for boosted black strings easier to understand than the Penrose process for boosted Schwarzschild black holes.

\begin{figure}
\includegraphics[width=\columnwidth]{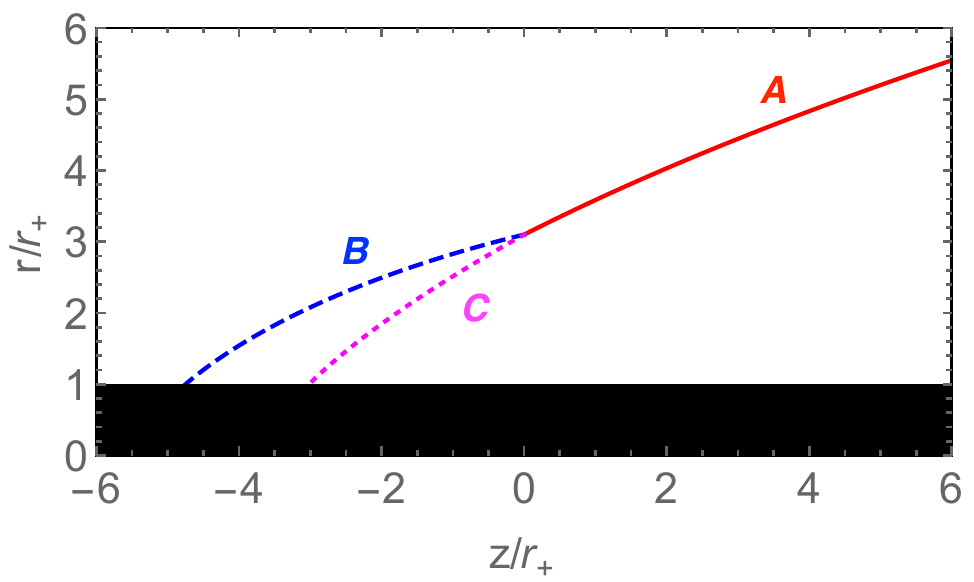}
\caption{Boosted black string with horizon at $r_+=1$ and momentum in the $+z$ direction.  Timelike particle $A$ (solid red) travels from infinity toward the black string along a geodesic.  Inside the ergosphere, it splits into massless daughter particles $B$ (dashed blue) and $C$ (dotted magenta), which also follow geodesics.  $C$ has negative energy. $B$ and $C$ both fall into the horizon. }
\label{fig:penrosestring}
\end{figure}

However, there do not appear to be geodesic Penrose process solutions in this spacetime for an entirely new reason.  Recent work  \cite{Brito:2015} has shown  that if such solutions exist, the interaction point cannot be a turning point of the incoming particle.  We numerically searched for solutions for which the interaction point is not a turning point but  were unable to find any examples.    Fig. \ref{fig:penrosestring} shows a typical failed solution.  Particles $A$, $B$, and $C$ all follow geodesics with constant energy and momentum along $z$. Particle $A$ enters the ergosphere of the boosted black string and splits in two.  Particle $C$ falls into the black string with negative energy.  However, particle $B$ also falls into the black string.  The horizon is infinitely extended and it is impossible for $B$ to travel around it along a geodesic.

\section{Jets}
\label{sec:jets}

The BZ model \cite{1977MNRAS.179..433B,1986bhmp.book.....T} is the electromagnetic younger cousin of the Penrose process.   It describes how electromagnetic fields can extract the rotational energy of Kerr black holes and it is widely believed to describe astrophysical jets.  In this section, we develop the analogue of the BZ jet power prediction for boosted Schwarzschild black holes.

\subsection{Coordinates}

The Schwarzschild metric in Schwarzschild coordinates, $(t,r,\theta,\phi)$, is
\beq\label{eq:gSchwarzschild}
ds^2 = -\left(1-\frac{2M}{r}\right)dt^2+\frac{dr^2}{1-2M/r}+r^2d\Omega^2,
\eeq
where $d\Omega^2=d\theta^2+\sin^2\theta d\phi^2$.
The fiducial observer (FIDO) frame is
\begin{align}
e^{\hat{t}} &= \sqrt{-g_{tt}} dt,\label{eq:zamot}\\
e^{\hat{r}} &= \sqrt{g_{rr}} dr,\label{eq:zamor}\\
e^{\hat{\theta}} &= \sqrt{g_{\theta\theta}} d\theta,\label{eq:zamoh}\\
e^{\hat{\phi}} &= \sqrt{g_{\phi\phi}} d\phi.\label{eq:zamop}
\end{align}
The relationship with KS coordinates \eqref{eq:gKS} is
\begin{align}
d\tau &= dt + \frac{2M}{r-2M} dr,\\
x&=r\sin\theta\cos\phi,\\
y&=r\sin\theta\sin\phi,\\
z&=r\cos\theta.
\end{align}
Define boosted Schwarzschild coordinates, $(t',r',\theta',\phi')$, by
\begin{align}
 dt'         &= d\tau'- \frac{2 M }{r-2 M} dr', \\
 dr'          &=  \sin \theta \cos \phi dx' + \sin \theta \sin \phi dy'+ \cos \theta dz', \\
 d\theta' &=  \frac{\cos \theta \cos \phi}{r} dx'+ \frac{\cos \theta \sin \phi}{r} dy' -\frac{\sin \theta}{r}dz', \\
 d\phi'    &=-\frac{\csc \theta \sin \phi}{r}dx' + \frac{\csc \theta\cos \phi }{r}dy'.
\end{align}
where $(\tau',x',y',z')$ are defined by \eqref{eq:tp}-\eqref{eq:yp}.  In primed coordinates the black hole carries momentum along $z$.  The reverse transformation is
\begin{align}
d\tau' &= dt' + \frac{2M}{r-2M} dr',\label{eq:dtaupblboost}\\
dx'&=\sin\theta\cos\phi dr'+r\cos\theta\cos\phi d\theta' \notag\\
&-r\sin\theta\sin\phi d\phi',\\
dy'&=\sin\theta\sin\phi dr'+r\cos\theta\sin\phi  d\theta'\notag\\
&+r\sin\theta\cos\phi d\phi',\\
dz'&=\cos\theta dr'-r\sin\theta d\theta'.
\end{align}

Useful transformations between these reference frames are collected in Appendix \ref{sec:frames}.

\subsection{Jet power}
\label{sec:pjetinf}

Define the jet power to be
\beq\label{eq:edotinf}
P_{\rm jet}  \equiv \frac{dE'}{dt'} = -\int_{S^2} T^{r'}_{t'} \sqrt{-g'} d\theta' d\phi',
\eeq
so $P_{\rm jet}>0$ corresponds to energy leaving the black hole.  The vector $\partial_{t'}$ is not Killing, so $P_{\rm jet}$ may be a function of radius.  We are interested in the jet power at infinity (the jet power at the horizon is computed Appendix \ref{sec:EMH}).  In our idealized setup, we assume an isolated black hole with a jet extending to infinity.  Astrophysical jets extend far beyond the horizon, so our idealized setup is a good approximation.

The FIDO-frame components of $T^r_{t'}$  are 
\beq\label{eq:trtinf}
T^r_{t'} = \gamma \left(\alpha^2+\frac{2M}{r}v\cos\theta\right)T^{\hat{r}}_{\hat{t}}
-\gamma v \cos\theta T^{\hat{r}}_{\hat{r}}
+\gamma v \alpha \sin \theta T^{\hat{r}}_{\hat{\theta}},
\eeq
where $\alpha^2=1-2M/r$. 
The advantage of the FIDO frame is that $T^{\hat{\mu}\hat{\nu}}$ is simply \cite{1973grav.book.....M}
\begin{align}
T^{\hat{t}\hat{t}} &= \frac{1}{2}(\mathbf{E}^2+\mathbf{B}^2),\label{eq:t00}\\
T^{\hat{t}\hat{j}} &= T^{\hat{j}\hat{t}} = (\mathbf{E}\times\mathbf{B})^{\hat{j}},\label{eq:t0j}\\ 
T^{\hat{j}\hat{k}} &= -E^{\hat{j}} E^{\hat{k}} - B^{\hat{j}} B^{\hat{k}}
+\frac{1}{2}(\mathbf{E}^2+\mathbf{B}^2) \delta^{\hat{j}\hat{k}},\label{eq:tjk}
\end{align}
where $\mathbf{E}$ and $\mathbf{B}$ are the FIDO-frame electric and magnetic fields.

At infinity, the six components of the electromagnetic field are not all independent because radiation is always outgoing at \ip.  In particular, in the large $r$ limit, we have the boundary condition
\beq\label{eq:bcinf}
\vec{E}_\parallel  = -n \times \vec{B}_\parallel,
\eeq
where $\vec{E}_\parallel = (E^{\hat{\theta}},E^{\hat{\phi}})$, $\vec{B}_\parallel = (B^{\hat{\theta}},B^{\hat{\phi}})$, and $n$ is the outward-pointing unit normal vector  (see Appendix \ref{sec:outgoing} for a derivation).
In components, 
\beq
E^{\hat{\theta}}=B^{\hat{\phi}},\quad E^{\hat{\phi}}=-B^{\hat{\theta}}.
\eeq
This eliminates two components of the fields at infinity.  

We further enforce the force-free constraint $\mathbf{E}\cdot\mathbf{B}=0$, which is a good approximation for astrophysical black hole magnetospheres \cite{1977MNRAS.179..433B,1986bhmp.book.....T}. In astrophysical jets, the force-free condition breaks down far from the black hole, in the so-called load region, where gas kinetic energy becomes comparable to the magnetic energy of the jet.  The force-free condition is a good approximation between the horizon and the load region.  The load is believed to be sufficiently far from the black hole so that for our purposes we may place it at infinity (see, e.g., \cite{2013MNRAS.436.3741P}). Combining the outgoing boundary condition with the force-free constraint implies  $E^{\hat{r}}=0$ or $B^{\hat{r}}=0$ at large $r$.    The outgoing boundary condition \eqref{eq:bcinf}  also implies $F^2=2(B^2-E^2)=2(B_{\hat{r}}^2-E_{\hat{r}}^2)$ at large $r$.      Astrophysical fluids are magnetically dominated because the electric field vanishes in the rest frame of highly ionized plasma.  So we choose $E^{\hat{r}}=0$ at large $r$.   This is the usual choice in astrophysics and it is the case that has been simulated (e.g., \cite{2011PNAS..10812641N}).

It is helpful to replace $\vec{E}_\parallel$ and $\vec{B}_\parallel$ with the field line velocity $\mathbf{v}_F$, defined by 
$\mathbf{E} = -\mathbf{v}_F\times \mathbf{B}$.  
In components, the fields at infinity become
\begin{align}
E^{\hat{\phi}} &= -B^{\hat{\theta}} = \frac{v^{\hat{\theta}}_F}{1-v^{\hat{r}}_F}B^{\hat{r}},\label{eq:epinf}\\
E^{\hat{\theta}} &= B^{\hat{\phi}} = -\frac{v_F^{\hat{\phi}}}{1-v_F^{\hat{r}}}B^{\hat{r}}.
\end{align}
Plugging into \eqref{eq:trtinf} gives the stress-energy tensor at infinity,
\beq\label{eq:trtinf2}
T^r_{t'}= -\gamma \left(\frac{v_F^\parallel}{1-\vfr}\right)^2 B_{\hat{r}}^2
  +\frac{1}{2} \gamma v \cos\theta B_{\hat{r}}^2
  +\gamma v \sin\theta \frac{v^{\hat{\theta}}_F}{1-v^{\hat{r}}_F} B_{\hat{r}}^2,
\eeq
where $v_F^\parallel\equiv\sqrt{(\vfh)^2+(\vfp)^2}$.

Assume small velocities: $v_F/r \sim v\ll 1$.  
A slowly moving black hole ($v\ll 1$) is assumed for simplicity. 
Small $v_F/r$ should be a good assumption in this case because we expect $v_F/r \sim v$ (just as in the BZ model for spinning black holes).  In this limit,
\beq\label{eq:trtsimple}
T^r_{t'}=- (v^\parallel_F)^2 B_{\hat{r}}^2
  +\frac{1}{2} v \cos\theta  B_{\hat{r}}^2
  +v v^{\hat{\theta}}_F  \sin\theta B_{\hat{r}}^2,
\eeq
and 
\beq\label{eq:pjetinf}
\frac{dE'}{dt'} =- \int_{S^2} T^r_{t'} \sqrt{-g}d\theta d\phi,
\eeq
so integrating \eqref{eq:trtsimple} over the sphere at infinity gives the jet power.
It depends on the unknown functions $v_F^\parallel$ and $B^{\hat{r}}$.  Unlike the original BZ model for spinning black holes, there are no exact force-free solutions to be our guide. There are less symmetries than in the BZ model, so it is unclear whether exact solutions are possible.    

For the moment, the best guide to $v_F^\parallel$ and $B^{\hat{r}}$ are numerical simulations.  Numerical simulations of BZ jets tend to relax to field geometries with $\Omega_F/\Omega_H\approx 1/2$ (where $\Omega_F$ and $\Omega_H$ are the field line and horizon angular velocities), and $B^{\hat{r}}$ is roughly uniform on the horizon (at least for low black hole spins) \cite{2004ApJ...611..977M,2013MNRAS.436.3741P}.  The field is approximately a split monopole.  The split monopole is in some sense the simplest solution and it acts like a ground state, while higher order multipoles are radiated away.  

We assume jets from boosted black holes are similar and guess
\beq\label{eq:target}
\frac{v^\parallel_F/r}{v/(2M)} = \frac{1}{2},
\eeq
and that $B^{\hat{r}}$ is a function of $r$ only.  
In this case, only the first term on the rhs of \eqref{eq:trtsimple} contributes to the integral \eqref{eq:pjetinf} and the jet power is
\beq\label{eq:pjet}
P_{\rm jet} = \frac{1}{4 \pi} \left(\frac{v}{2M}\right)^2{\Phi}^2,
\eeq
where $\Phi= (2\pi r^2 B^{\hat{r}})_{r\rightarrow \infty}$ is the flux through a hemisphere at infinity.  This is similar to the BZ prediction, $P_{\rm jet}^{\rm BZ} = \Omega_H^2 \Phi_H^2/(6\pi)$, but with $v/(2M)$ playing the role of $\Omega_H$, the flux measured at infinity rather than the horizon, and a slightly different normalization constant.  The upshot is that boosted black holes and spinning black holes have jets of comparable strength when $v/(2M)\sim \Omega_H$ (for fixed magnetic flux).   

The jet power observed in numerical simulations of boosted black holes appears to be smaller than \eqref{eq:pjet} by as much as a factor of 100 \cite{2011PNAS..10812641N}.  It may be that \eqref{eq:target} is an overestimate of the field line velocity.  It may also be relevant that the simulated jets do not extend over a full $4\pi$ steradians.  It will be interesting to understand this difference better but we save a more detailed comparison for the future.  

The membrane paradigm gives a dual description of black holes as conductive membranes \cite{1986bhmp.book.....T,2013MNRAS.436.3741P}.  
The power radiated by a conductor moving through a magnetic field scales with velocity and field strength as $P\sim v^2 B^2$ \cite{1965PhRvL..14..171D,1965JGR....70.3131D}.  
 So the black hole jet power may also be expected to scale as $v^2B^2$ \cite{2011PNAS..10812641N}.  The jet power formula \eqref{eq:pjet} confirms this expectation.   The power radiated by a conductor scales with the size of the conductor as $L^2$  \cite{1965PhRvL..14..171D,1965JGR....70.3131D}. This shows up in our formula as a factor of $M^2$.
 
\section{The membrane at infinity}
\label{sec:membrane}

The BZ model has an elegant formulation in the black hole membrane paradigm \cite{1986bhmp.book.....T,2013MNRAS.436.3741P}.  In this picture, the black hole is represented by a fluid membrane at the horizon.  The black hole's mass and angular momentum are stored in the membrane's stress-energy tensor and jets are powered by electromagnetic torques acting on the membrane.

In our model of jets from boosted black holes, the energy flux at the horizon need not match the energy flux at infinity because $\partial_{t'}$ is not Killing.   So in this section we will reformulate the membrane paradigm such that the membrane lives at infinity (where the jet power is evaluated) rather than the horizon.

We begin by reviewing the standard black hole membrane paradigm.  A modern derivation is based on an action principle \cite{Parikh:1997ma}.  Consider an observer who remains forever in the black hole exterior.  Such an observer cannot receive signals from the black hole interior, so the interior can be eliminated from their calculations.  In particular, given a Lagrangian, $\mathcal{L}$, they can use the action
\beq
S=\int_{\rm exterior} d^4x \sqrt{-g} \mathcal{L},
\eeq   
with domain of integration restricted to the black hole exterior.  The variation of this action, $\delta S$, gives boundary terms supported on the horizon.  To obtain the correct equations of motion, the boundary terms need to be eliminated by adding surface terms to the action.  The surface terms encode the properties of the membrane on the horizon.   In particular, they fix the membrane's current density and stress-energy tensor.  Further imposing the boundary condition that all waves are ingoing at the horizon fixes the resistivity and viscosity of the membrane.   

The true horizon is a null surface.  It is convenient to define the membrane on a stretched horizon, a timelike surface some small distance above the true horizon, and then take the true horizon limit.

\begin{figure}
\includegraphics[width=\columnwidth]{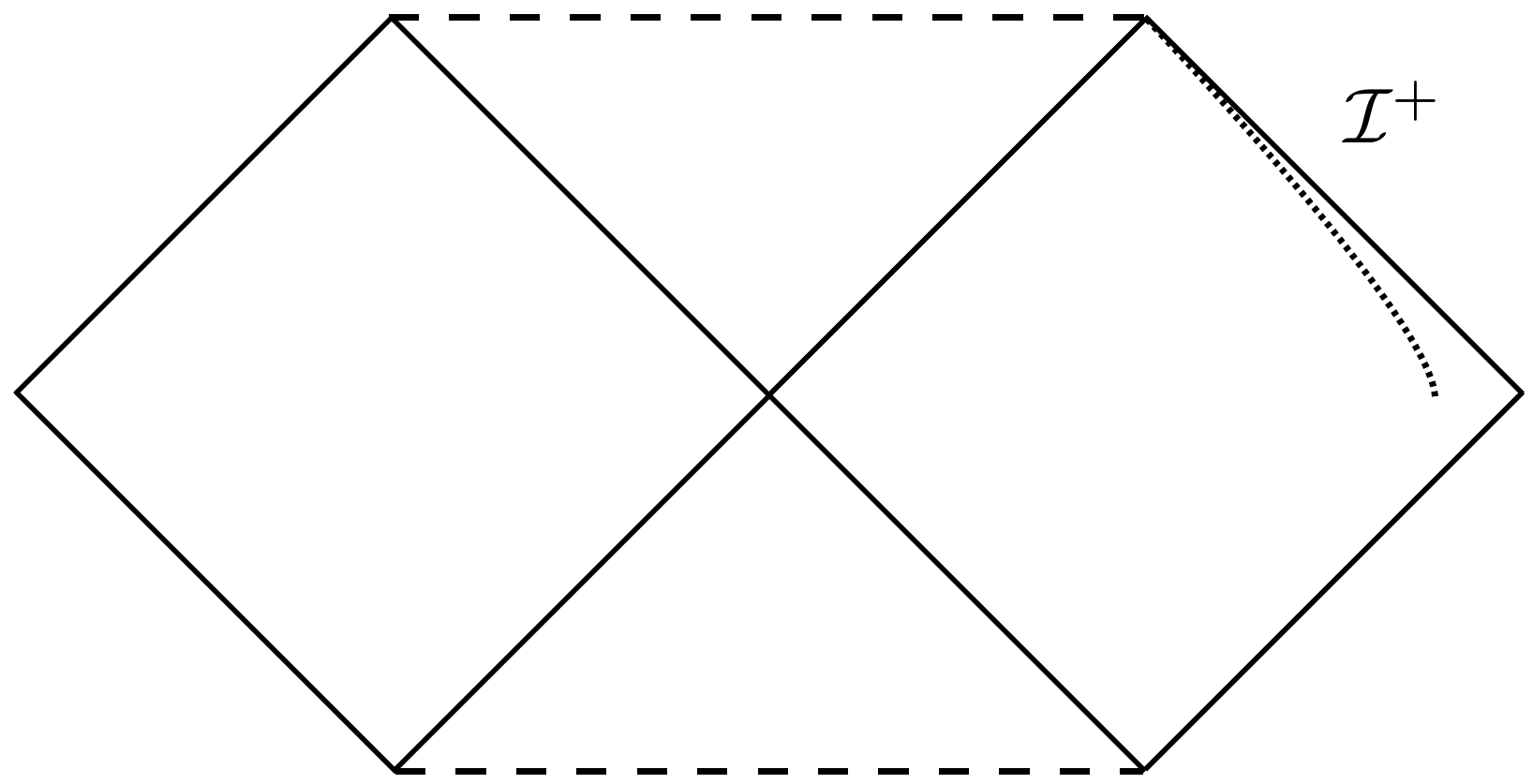}
\caption{Black hole Penrose diagram.  Stretched infinity (dotted) is a timelike surface some large but finite distance from the black hole. }
\label{fig:infinity}
\end{figure}

This section is based on the observation that the same recipe works at \ip.  Consider an observer who remains forever in the black hole exterior.  They cannot receive signals from beyond \ip.  Define ``stretched infinity'' to be a timelike surface some large but finite distance from the black hole (see Fig. \ref{fig:infinity}).  Let  $\mathcal{M}'$ be a truncated spacetime ending at stretched infinity.  Given a Lagrangian $\mathcal{L}$, use the action
\beq
S= \int_{\mathcal{M}'} d^4x \sqrt{-g} \mathcal{L},
\eeq
with domain of integration $\mathcal{M}'$.
Varying this action gives boundary terms supported on stretched infinity, which must be canceled by adding surface terms to the action.  These surface terms fix the current and stress-energy tensor of the membrane at infinity.  The boundary condition that all waves are outgoing at \ip\ fixes the resistivity and viscosity of the membrane.

\subsection{Membrane current}

To derive the electromagnetic properties of the membrane at infinity, consider the Maxwell action 
\beq\label{eq:maxwell}
S = \int d^4x \sqrt{-g} \left(-\frac{1}{4}F^2+J\cdot A\right).
\eeq
Varying this action gives a term which is a total derivative
\beq\label{eq:totald}
-\int \partial_a (\sqrt{-g}  F^{ab}\delta A_b)d^4x,
\eeq
and integrating by parts gives a surface term supported on stretched infinity,
\beq\label{eq:surfinf}
-\int d^3 x \sqrt{-h} F^{ab}n_a \delta A_b,
\eeq
where $h$ is the determinant of the induced metric on stretched infinity and $n^a$ is the outward-pointing spacelike unit normal at stretched infinity.  We conclude that stretched infinity carries a current
\beq\label{eq:jinf}
j^a = -F^{ab}n_b.
\eeq 
Its time component,
\beq
\sigma = -F^{tb}n_b = -E_\perp,
\eeq
is the membrane's charge density and it terminates the normal component of the electric field at stretched infinity.  The spatial components of $j^a$ form a surface current terminating the tangential components of the magnetic field,
\beq\label{eq:membranej}
\vec{B}_\parallel = \hat{n} \times \vec{j}.
\eeq

The current \eqref{eq:jinf} contains an overall minus sign relative to the current in the usual black hole membrane paradigm \cite{1986bhmp.book.....T,Parikh:1997ma}.  This may be traced to the integration by parts of \eqref{eq:totald} and the fact that stretched infinity is an outer boundary of $\mathcal{M}'$ whereas the horizon is an inner boundary.  The minus sign has a simple physical interpretation: outward-pointing radial field lines begin at positive charges on the stretched horizon and terminate at negative charges on stretched infinity.  The charge density of stretched infinity vanishes in the true infinity limit.  However, the surface area blows up in this limit, so the total charge of the membrane at infinity remains finite.

At stretched infinity, we have the outgoing boundary condition \eqref{eq:bcinf}.  Combined with \eqref{eq:jinf}, it implies Ohm's law,
\beq\label{eq:ohm}
 \vec{E}_\parallel = \rho \vec{j},
\eeq
on the membrane at infinity.  Equations  \eqref{eq:bcinf} and \eqref{eq:jinf} at stretched infinity differ from the   black hole horizon versions by relative minus signs, but these signs cancel in \eqref{eq:ohm}. So the resistivity of the membrane at infinity has the same value as in the usual black hole membrane paradigm, $\rho = 1 = 377 \Omega$.

\subsection{Membrane stress-energy tensor}

Now consider the Einstein-Hilbert action.  Varying the action on $\mathcal{M}'$ gives a surface term supported on stretched infinity.  Eliminating this surface term endows stretched infinity with a stress-energy tensor
\beq\label{eq:tab}
t_{ab} = -\frac{1}{8\pi}(K h_{ab}-K_{ab}),
\eeq
where $h_{ab}$ is the induced metric on stretched infinity,
\beq\label{eq:K}
K^a_b = {n^a}_{|b},
\eeq
is its extrinsic curvature, $K={K^a}_a$, and $_{|b}$ is the three-covariant derivative on stretched infinity.  This is the same stress-energy tensor that appears in the original black hole membrane paradigm \cite{1986bhmp.book.....T,Parikh:1997ma} but with an overall minus sign.  As in the previous section, the sign comes from the fact that $t_{ab}$ is obtained from an integration by parts and stretched infinity is an outer boundary of spacetime.   Equation \eqref{eq:tab} is the same as the Brown-York stress-energy tensor but with an overall minus sign.  We explain the origin of this difference below.

Just as the membrane's current terminates electric and magnetic fields, the membrane's stress-energy tensor creates a discontinuity in the extrinsic curvature.  The discontinuity is given by the Israel junction condition \cite{1986bhmp.book.....T,Parikh:1997ma}
\beq\label{eq:israel}
t_{ab} = \frac{1}{8\pi} ([K]h_{ab}-[K]_{ab}),
\eeq
where $[K]=K_+-K_-$ is the difference between the extrinsic curvature of stretched infinity as defined with respect to the spacetime outside stretched infinity and as defined with respect to the spacetime inside.  The extrinsic curvature appearing in \eqref{eq:tab} is $K_-$, so the Israel junction implies $K_+=0$. In other words, the membrane stress-energy tensor \eqref{eq:tab} terminates the gravitational field outside stretched infinity.  The Brown-York stress-energy tensor is defined so as to terminate the gravitational field inside \ip.  This explains the relative minus sign between \eqref{eq:tab} and the Brown-York stress-energy tensor.

The analogue of the electromagnetic outgoing boundary condition \eqref{eq:bcinf} is encoded in the relationship between the extrinsic curvature of stretched infinity, ${K^a}_b$, and the extrinsic curvature of true infinity, 
\beq\label{eq:knull}
{k^a}_b = {l^a}_{|b},
\eeq
where $l$ is the future-directed null generator of \ip.   Null generators are normal to true infinity because it is a null surface, and the future-directed null generator plays the role of the outward-pointing normal in the definition of extrinsic curvature for null surfaces.  

As stretched infinity approaches true infinity, 
\beq\label{eq:nlimit}
n^a \rightarrow -l^a,
\eeq
and so
\beq\label{eq:klimit}
{K^a}_b \rightarrow -{k^a}_b.
\eeq
The minus sign reflects the fact that all radiation at \ip\ is outgoing.  At a black hole horizon the sign would be positive.

To summarize, the membrane at infinity differs from the membrane at the horizon by two extra minus signs.  The first minus sign is the overall sign in \eqref{eq:tab}.  This minus sign appears because \ip\ is an outer boundary of spacetime rather than an inner boundary.  The second extra minus sign is the sign in \eqref{eq:nlimit}.  This minus sign appears because \ip\ satisfies an outgoing rather than an ingoing boundary condition.  These two minus signs are independent.  For example, at \im\ only the first extra minus sign would appear. At a white hole horizon only the second extra minus sign would appear.

To clarify the minus sign in \eqref{eq:nlimit}, consider the Schwarzschild spacetime \eqref{eq:gSchwarzschild}.    Ingoing and outgoing Eddington-Finkelstein coordinates are 
\begin{align}
v&=t+r^*\label{eq:ef1}\\
u&=t-r^*,\label{eq:ef2}
\end{align}
where
\beq
\frac{dr^*}{dr}=\left(1-\frac{2M}{r}\right)^{-1}.
\eeq
Stretched infinity is a timelike surface at some large but finite radius.  Its outward-pointing unit spacelike normal is
\beq
\hat{n} = \partial_{r*}.
\eeq
The future-directed null generator of true infinity is 
\beq
l = 2 \partial_{u},
\eeq  
where the normalization is a convention that leads to simpler formulas.  On \ip, $v={\rm const}$ and $dv=0$.  In this case, \eqref{eq:ef1}-\eqref{eq:ef2} give 
\beq
du=2dt = -2dr_*,
\eeq 
and so,
\beq
\hat{n}=\partial_{r*}=-\partial_{t}=-2\partial_u =-l.
\eeq
As stretched infinity approaches true infinity, $\hat{n}\rightarrow -l$, as claimed.  This explains the minus sign in \eqref{eq:klimit}.

Enforcing the boundary condition \eqref{eq:klimit} turns $t_{ab}$ into the stress-energy tensor of a viscous fluid.  Split spacetime into space and time by fixing a family of fiducial observers with four-velocity $U^a$ such that $U^a\rightarrow l^a$ at true infinity.  (For Schwarzschild, these are the FIDOs.)  Define constant-time surfaces to be surfaces to which $U^a$ is orthogonal.
The metric on a two-dimensional constant-time slice of stretched infinity is
\beq
\gamma_{AB} = h_{AB} + U_A U_B,
\eeq
where uppercase indices $A, B,\dots$ indicate tensors living on these slices.

The time-time component of the extrinsic curvature  is
\beq
U^a U_b {k^b}_a= -\kappa,
\eeq
where the surface gravity, $\kappa$, is defined by $l^a\nabla_a l^b=\kappa l^b$, and we have used Eqs. \eqref{eq:knull} and \eqref{eq:nlimit}.
Decompose the space-space components of the extrinsic curvature  into a traceless part and a trace,
\beq
k_{AB} = \sigma_{AB} + \frac{1}{2}\gamma_{AB}\theta,
\eeq
where $\sigma_{AB}$ is the shear and $\theta$ the expansion.  The time-space components vanish: $U^b k^A_b = 0.$  The trace is $k={k^A}_A=\kappa+\theta$.

Plugging into \eqref{eq:tab} gives the stress tensor of the membrane at infinity,
\beq\label{eq:tab2}
t_{AB} = \frac{1}{8\pi}\left(-\sigma_{AB}+\gamma_{AB}\left(\frac{1}{2}\theta+\kappa\right)\right).
\eeq
It is the usual stress tensor of a two-dimensional viscous Newtonian fluid with pressure $p=\kappa/(8\pi)$, shear viscosity $\eta=1/(16\pi)$, and bulk viscosity $\zeta = -1/(16\pi)$.  Equations \eqref{eq:tab} and \eqref{eq:klimit} differ from the stretched horizon versions by relative minus signs but these signs cancel in \eqref{eq:tab2}, so the viscosity parameters of the membrane at infinity are the same as in the standard membrane paradigm at the black hole horizon.  

\subsection{Jets revisited}

Consider the momentum flux,
\beq
\frac{dP}{dt'} = \int T^r_{z'}\sqrt{-g} d\theta d\phi,
\eeq
at stretched infinity for a boosted Schwarzschild black hole.  For small $v$, the only contribution is the term
\beq
T^r_{z'} = -\sin\theta {T^{\hat{r}}}_{\hat{\theta}}  = \sin\theta B^{\hat{r}}B^{\hat{\theta}}.
\eeq
In membrane variables, the momentum flux is,
\beq\label{eq:pdotmembrane}
\frac{dP}{dt'} = \int (\vec{j}\times\vec{B})^z \sqrt{-g}d\theta d\phi,
\eeq
where we have used  \eqref{eq:membranej}.  This is the usual expression for a Lorentz force acting on the membrane at infinity.

For small $v$, the energy flux at infinity is
\beq
\frac{dE'}{dt'} 
= \int (E^{\hat{\theta}}B^{\hat{\phi}}-E^{\hat{\phi}}B^{\hat{\theta}})\sqrt{-g}d\theta d\phi.
\eeq
Using the outgoing boundary condition \eqref{eq:bcinf} and Ohm's law \eqref{eq:ohm} gives
\beq\label{eq:edotmembrane}
\frac{dE'}{dt'}
=\int \rho |\vec{j}|^2\sqrt{-g}d\theta d\phi,
\eeq
the usual expression for Joule heating in a resistor.

\subsection{Dual current formulation}

The membrane current, $j^a$, encodes all components of the electromagnetic field at infinity except $B^{\hat{r}}$.  There is an alternate formulation of membrane electrodynamics in which all the variables we need at infinity are components of the membrane current.
Start not from the usual Maxwell action \eqref{eq:maxwell}, but rather
\beq
S=-\frac{1}{4}\int (*F)^2 \sqrt{-g} d^4x,
\eeq
where $*F$ is the dual field strength.
Then the membrane's current density is
\beq
j^a_* = -*\! F^{ab}n_b,
\eeq
instead of \eqref{eq:jinf}.
It is a magnetic monopole current.  The magnetic monopole charge density is
\beq
\sigma_* \equiv j^{\hat{t}}_* = -B^{\hat{r}},
\eeq
and it terminates the normal component of the magnetic field. 
The idea of terminating the magnetic field at the horizon with monopole charges has been suggested by \cite{1978MNRAS.185..833Z}. 
The other components of the monopole current are
\begin{align}
j^{\hat{\theta}}_* &= E^{\hat{\phi}} = -B^{\hat{\theta}},\\
j^{\hat{\phi}}_* &= - E^{\hat{\theta}} = -B^{\hat{\phi}}.
\end{align}
The only component of the field not packaged in $j^{\hat{i}}_*$ is $E^{\hat{r}}$, but force-free jets have $E^{\hat{r}}=0$ at stretched infinity.  So $j^{\hat{i}}_*$ includes all the electromagnetic degrees of freedom we need at infinity.  

In these variables, the momentum flux \eqref{eq:pdotmembrane} is
\beq
\frac{dP}{dt'} = \int \sigma_* B^z \sqrt{-g}d\theta d\phi,
\eeq
which is the magnetic monopole equivalent of a $q\mathbf{E}$ Lorentz force. The torques driving standard BZ jets are $\sigma_* B^\phi$ Lorentz forces.  The energy flux is the same as \eqref{eq:edotmembrane} but with $|\vec{j}_*|^2$ in place of $|\vec{j}|^2$.  The advantage of the dual current formulation is that all the variables at infinity live in 2+1 dimensions.

\section{Conclusions}
\label{sec:conc}

We have developed the theory underlying kinetic energy extraction from moving black holes.  We derived the analogues of the Penrose process and the BZ jet power prediction for boosted black holes.  We also derived a new version of the membrane paradigm in which the membrane lives at infinity, and we showed that this formalism is useful for interpreting energy extraction from boosted black holes.   

The Penrose processes for boosted black holes and spinning black holes have a similar conceptual basis.  In both cases, energy extraction is related to the existence of negative energy trajectories.    BZ jets are a generalized version of the Penrose process, with force-free electromagnetic fields replacing point particles.  So jets from boosted black holes and spinning black holes are also qualitatively similar.  The same language that describes jets from spinning black holes (e.g., negative energy fluxes inside the ergosphere, torques acting on a membrane) can be applied to jets from boosted black holes.

We have highlighted two important technical differences between boosted black holes and spinning black holes.  One is that the relevant notion of energy in the boosted case is defined with respect to a vector $\partial_{t'}$ which is not Killing.  As a result, the energy flux at the horizon need not match the energy flux at infinity even for $\partial_{t'}-$invariant solutions.   One can construct solutions in which the energy fluxes at the horizon and infinity are the same (as we showed in Sec. \ref{sec:penrose}), but astrophysically relevant solutions (such as the jets in Sec. \ref{sec:jets}), are unlikely to have this property.  So it is important to compute fluxes at infinity.  

A second difference between energy extraction from boosted black holes and spinning black holes is that the former is an observer-dependent process, while the latter is observer independent.  This can be traced to the fact that the relevant notion of black hole energy in the boosted case is the time component of $P^{\rm ADM}_\mu$, which is not a Lorentz invariant.  The relevant notion of black hole energy in the spinning case is the norm $-P_{\rm ADM}^2$, which is Lorentz invariant.

Our discussion of the Penrose process for boosted black holes in Sec. \ref{sec:penrose} relied on  numerical solutions for trajectories with constant linear momentum.  It may be possible to find and classify these trajectories analytically.  This would allow one to answer a number of interesting questions.  For example, what is the maximum energy that can be extracted using the boosted black hole Penrose process as a function of the interaction point $(r_*,\theta_*)$?  The answers are somewhat coordinate dependent, but understanding the answers in a natural coordinate system would give insight into general features of the process. 

We have described the analogue of BZ jets and computed the jet power \eqref{eq:pjet} to be
\beq
P_{\rm jet} = \frac{1}{4 \pi} \left(\frac{v}{2M}\right)^2{\Phi}^2,
\eeq
at least for small $v$.  This can be tested with numerical simulations \cite{2010Sci...329..927P,2011PNAS..10812641N}.   It will be interesting to use simulations to understand the distributions of $\Omega_F$ and $B^{\hat{r}}$ and to compare the energy and momentum fluxes at the horizon and infinity.  On the analytical side, our computations can be generalized away from the small $v$ limit and they can be generalized from boosted Schwarzschild black holes to boosted Kerr black holes.

We have shown that it is possible to reformulate the standard membrane paradigm such that the membrane lives at infinity rather than the black hole horizon.  The membrane at infinity has the same resistivity and viscosity coefficients as in the standard membrane paradigm.  The membrane at infinity is useful for understanding jets from boosted black holes because the energy and momentum fluxes at infinity can be described using the familiar language of dissipation and Lorentz forces acting on a conductor.  

The stress-energy tensor of the membrane at infinity is the same as the Brown-York stress-energy tensor \cite{Brown:1992br} up to a minus sign.  The Brown-York stress-energy tensor is not finite for general asymptotically flat spacetimes but requires the addition of Mann-Marolf counterterms \cite{Mann:2005yr}.  Similar counterterms should be incorporated into the definition of the membrane at infinity.  We hope to explore the membrane interpretation of these counterterms in the future.

\begin{acknowledgments}
I thank Richard Brito, Vitor Cardoso, Carlos Palenzuela, Paolo Pani, Maria Rodriguez and especially Luis Lehner for helpful comments and suggestions.  This work was supported by a Pappalardo Fellowship in Physics at MIT.

\end{acknowledgments}

\appendix

\section{Boost invariance of horizon area}
\label{sec:boostinv}

The discussion of Sec. \ref{sec:adm} relied on the fact that the area of a Schwarzschild black hole's event horizon is boost invariant.  This follows from the more general fact that the area of an event horizon with vanishing expansion is slicing invariant.  This is a well-known statement (see e.g. \cite{2010GReGr..42..387A}) but we record a proof here for completeness.

Consider a foliation of spacetime into spacelike slices, $\Sigma$, with future-pointing unit normal $\xi^a$.  The horizon is a 2-sphere in $\Sigma$ with outward-pointing unit normal $n^a$.  The induced metric on the horizon is
\beq
m_{ab}=g_{ab} + \xi_a \xi_b - n_a n_b.
\eeq
The presence of $\xi_a$ in this formula suggests the area computed using $m_{ab}$ might be slicing dependent.  Let
\beq
 \xi_a \xi_b - n_a n_b = 2 k_{(a}^+ k_{b)}^-,
\eeq
where $k_a^{\pm} = (\xi_a \pm n_a)/\sqrt{2}$ are null vectors.  
We can rescale $k^\pm$ such that $k^+=l$ and $k^+_a k^a_- = -1$.  

The area of the horizon is 
\beq
\mathcal{A} = \int_{S^2} m^{1/2} d^2 x,
\eeq
where $m=\det(m_{ab})$.  The choice of $S^2$ depends on the slicing, but one $S^2$ can be carried into another by translations along $l$. For the area to be slicing invariant, we require
\beq
\mathcal{L}_l m^{1/2} = 0,
\eeq
which is equivalent to vanishing expansion:
\beq\label{eq:expansion}
\theta = \nabla_a l^a = \frac{1}{m^{1/2}} \mathcal{L}_l m^{1/2}=0.
\eeq 

The second equality in \eqref{eq:expansion} follows from the Jacobi formula for Lie derivatives.
Let $a_{ij}$ be a nonsingular matrix and let $a=\det a_{ij}$.
Then the Jacobi formula is
\beq\label{eq:jacobi}
\mathcal{L}_X a = a a^{ji} \mathcal{L}_X a_{ij}.
\eeq
To prove this formula, note that the determinant is a polynomial in the $a_{ij}$ such that there is the chain rule
\beq
\mathcal{L}_X a = \frac{\partial a}{\partial a_{ij}} \mathcal{L}_X a_{ij}.
\eeq
Replacing the partial derivatives with $a a^{ji}$ gives the Jacobi formula.  An application of the Jacobi formula gives
\beq\label{eq:expansionidentity}
\mathcal{L}_l \log m^{1/2} 
= \frac{1}{2}m^{ba}(l^c m_{ab;c}+m_{cb}l^c_{;a}+m_{ac}l^c_{;b})
= \nabla_a l^a,
\eeq
which is \eqref{eq:expansion}.  

\section{Penrose process solutions}
\label{sec:penroseapp}

In this section we detail the numerical method used to find the Penrose process solutions discussed in Sec. \ref{sec:penrose} and we give another example of such a solution.

Our task is to find three trajectories, $A$, $B$, and $C$, which meet at an interaction point $(r_*,\theta_*)$ such that four-momentum is conserved,
\beq\label{eq:econapp}
u_\mu^A = u_\mu^B + u_\mu^C \quad \text{at $(r_*,\theta_*)$}.
\eeq
We further require that $A$ and $B$ extend to infinity and $C$ falls into the black hole with negative energy in the boosted KS frame.   

We assume $A$ is timelike and $B$ and $C$ are null.  Each trajectory is then fully characterized by two constants, $u_\tau$ and $u_z$ (we set $u_y=0$).
Given these constants, a trajectory $(r(\tau),\theta(\tau))$ is fixed by the differential equations
\begin{align}
u_z &= -\frac{2M\cos\theta}{\alpha^2r} u_\tau +  \frac{\cos\theta}{\alpha^2} \dot{r}-r\sin\theta \dot{\theta},\label{eq:diffeq1}\\
m^2 &= -\frac{u_\tau^2}{\alpha^2} + \frac{\dot{r}^2}{\alpha^2}+r^2\dot{\theta}^2,\label{eq:diffeq2}
\end{align}
where $\alpha=\sqrt{1-2M/r}$.  Four-momenta in KS and Schwarzschild coordinates are related by
\begin{align}
u_\tau &= u_t,\label{eq:utauapp}\\
u_x &= -\frac{2M\sin\theta}{\alpha^2 r}u_t+\sin\theta u_r + \frac{\cos\theta}{r}u_\theta,\\
u_z &= -\frac{2M\cos\theta}{\alpha^2 r}u_t+\cos\theta u_r - \frac{\sin\theta}{r}u_\theta.\label{eq:uzapp}
\end{align}

We begin by fixing the interaction point $(r_*,\theta_*)$ and the two constants $u_\tau^C$ and $u_z^C$ that define particle $C$.  In Sec. \ref{sec:penrose}, we picked $(r_*,\theta_*)=(4M,\pi/8)$, $u_\tau^C=-1/5+\epsilon$, and $u_z^C=-1/5$, where $\epsilon=10^{-4}$.  

Next, we choose $u_t^A$.  In our example, $u_t^A=-3/2$ .  The remaining components of particle $A$'s four-momentum, $u_r^A$ and $u_\theta^A$,  are fixed by
\begin{align}
-1 &= u_A^\mu u^A_\mu =  -\frac{(u_t^A)^2}{\alpha^2} + \alpha^2 (u_r^A)^2 + \frac{(u_\theta^A)^2}{r^2}, \\
-1/2 &= u_A^\mu u^C_\mu = -\frac{u_t^A u_t^C}{\alpha^2} + \alpha^2 u_r^A u_r^C + \frac{u_\theta^Au_\theta^C}{r^2}, \label{eq:constraint2}
\end{align}
at the interaction point.  Equation \eqref{eq:constraint2} follows from energy conservation: $0=u_B^\mu u^B_\mu = -1- 2 u_A^\mu u^C_\mu$, and so $u_A^\mu u^C_\mu=-1/2$.
The four-momentum of particle $A$ in KS coordinates is given by \eqref{eq:utauapp}-\eqref{eq:uzapp}.

Finally, we fix the four-momentum of particle $B$ using energy conservation \eqref{eq:econapp}.  In particular,
\begin{align}
u_t^B &= u_t^A - u_t^C,\\
u_z^B &= u_z^A - u_z^C.
\end{align}
The trajectories of $A$, $B$, and $C$ are now fully determined.  We used trial and error to find $u_\tau^C$, $u_z^C$, and $u_\tau^A$ such that the trajectories of $A$ and $B$ extend to infinity and particle $C$ falls into the black hole.

\begin{figure}
\includegraphics[width=\columnwidth]{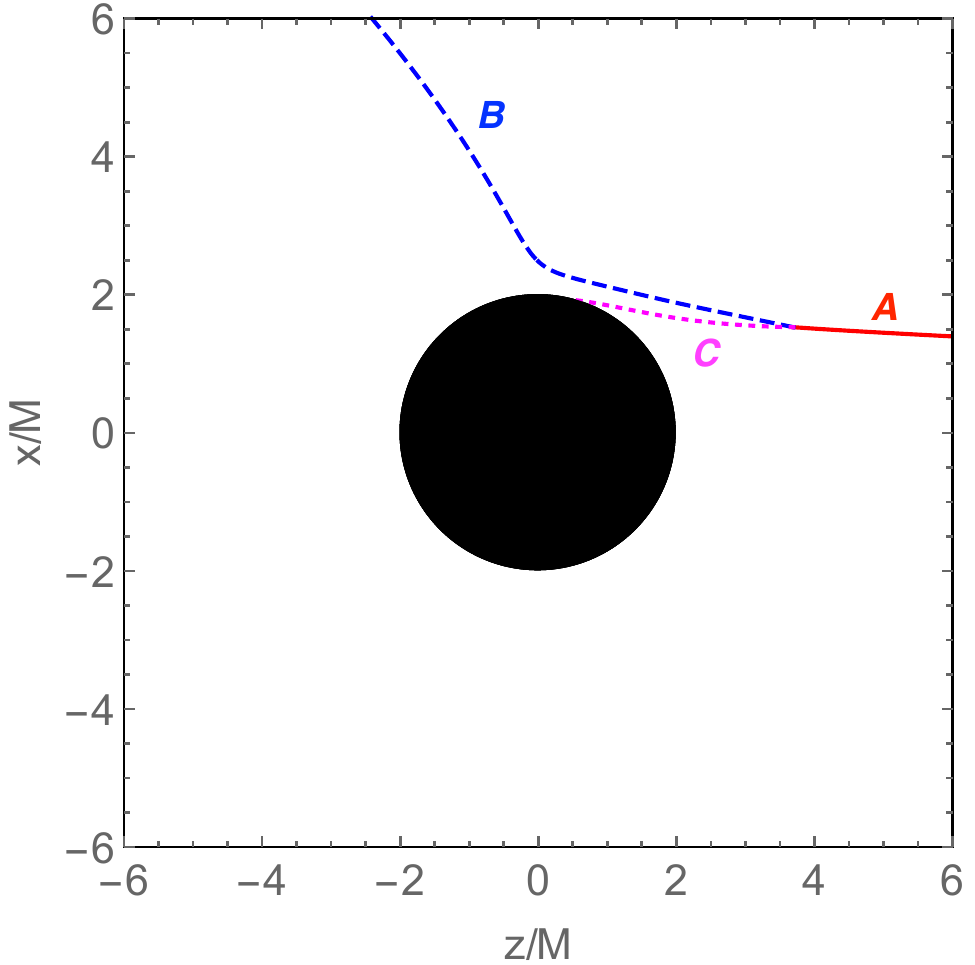}
\caption{Penrose process for a boosted Schwarzschild black hole moving in the $+z$ direction with velocity $v$.  This is similar to Fig. \ref{fig:penrose}, except the momentum of particle $A$ is primarily along $z$ (rather than $x$) and a larger amount of energy is extracted (for fixed $v$). }
\label{fig:penrose2}
\end{figure}

Fig. \ref{fig:penrose2} shows one such solution.  The interaction point is the same as in Sec. \ref{sec:penrose}, $(r_*,\theta_*)=(4M,\pi/8)$, but the momentum of $A$ is primarily along $z$ rather than $x$.  Particle $A$ has $(u_\tau^A,u_z^A)=(-10,-9.9376)$, particle $C$ has $(u_\tau^C,u_z^C)=(-6.99,-7)$, and the momentum of particle $B$ is fixed by energy conservation.  The energy of particle $C$ in the boosted frame is
\beq
E'_C = -\gamma(0.01-7(1-v)),
\eeq
which is negative for $v$ near $1$.

For this process to make sense, it is important that $u_z$ is finite at the horizon.  
A coordinate transformation gives
\beq\label{eq:danger}
u_z = -\frac{2M\cos\theta}{\alpha^2r}u_t+\cos\theta u_r-\frac{\sin\theta}{r}u_\theta. 
\eeq
The first two terms on the rhs are infinite at the horizon.  We need to check that these infinities cancel.  
Let us check this for a radial null geodesic (the general case is not much harder).  In this case,
\beq
0=u_\mu u^\mu  = -g^{tt}{u_t}^2 + g^{rr}(u_r)^2.
\eeq 
It follows that $u_r=g_{rr}u_t=u_t/\alpha^2$.  Plugging into \eqref{eq:danger} and setting $r=2M$ gives
\beq
u_z=\frac{\sin\theta}{r}u_\theta,
\eeq
which is finite.  

\section{Reference frames}
\label{sec:frames}

The discussion in Sec. \ref{sec:jets} relied on several different reference frames.  Here we collect some of the relevant transformations.  

The FIDO-frame components of the boosted Schwarzschild basis vectors are
\begin{align}
\partial_{t'}  &= 
 \gamma\left(\alpha  +\frac{2M  v \cos \theta}{\alpha  r}\right) e_{\hat{t}}
 -\frac{\gamma  v \cos \theta}{\alpha } e_{\hat{r}}\notag\\
 &+\gamma  v \sin \theta e_{\hat{\theta}},\label{eq:tpfido}\\
\partial_{r'} &=
\frac{2M \alpha ^2 (\gamma -1) \sin ^2\theta-\gamma  (r-4M) v \cos \theta}{\alpha ^3 r} e_{\hat{t}}\notag\\
&+\frac{1}{\alpha}\left(\sin ^2\theta+\gamma  \cos ^2\theta-\frac{2M \gamma  v \cos \theta}{\alpha^2 r}\right) e_{\hat{r}}\notag\\
&+\sin \theta \left(\cos \theta(1-\gamma)+\frac{2 M\gamma  v}{\alpha ^2 r}\right) e_{\hat{\theta}},\\
\partial_{\theta'} &= 
 \frac{\sin \theta \left(2M (\gamma -1) \cos \theta+\alpha ^2 \gamma  r v\right)}{\alpha } e_{\hat{t}}\notag\\
& -\frac{(\gamma -1) r \sin \theta \cos \theta}{\alpha } e_{\hat{r}}\notag\\
&+r \left(\gamma  \sin ^2\theta+\cos ^2\theta\right) e_{\hat{\theta}},\\
\partial_{\phi'} &= r \sin\theta e_{\hat{\phi}}.
\end{align}
The FIDO-frame components of the Schwarzschild one-forms are
\begin{align}
dt &= \frac{1}{\alpha} e^{\hat{t}},\\
dr &= \alpha e^{\hat{r}},\label{eq:drfido}\\
d\theta &= \frac{1}{r} e^{\hat{\theta}} ,\\
d\phi &= \frac{1}{r\sin\theta}e^{\hat{\phi}}.
\end{align}
Equation \eqref{eq:trtinf} follows from \eqref{eq:tpfido} and \eqref{eq:drfido}.

The boosted Schwarzschild components of the Schwarzschild one-forms are
\begin{align}
dt &= 
 \gamma\left(1 +\frac{2  M v \cos \theta}{\alpha ^2 r}\right) dt'\notag\\
&+\frac{2 \alpha ^2 (\gamma -1) M \sin ^2\theta+\gamma  v \cos \theta (4 M-r)}{\alpha ^4 r} dr'\notag\\
&+\frac{\sin \theta \left(2 (\gamma -1) M \cos \theta+\alpha ^2 \gamma  r v\right)}{\alpha ^2} d\theta', \label{eq:dtblboost}\\
dr &= 
- \gamma  v \cos \theta dt' \notag\\
&+ \left(\gamma  \cos ^2\theta+\sin ^2\theta-\frac{2 \gamma  M v \cos \theta}{\alpha ^2 r}\right)dr' \notag\\
&- (\gamma -1) r \sin \theta \cos \theta d\theta', \\
d\theta &=
 \frac{\gamma  v \sin \theta}{r} dt'\notag\\
 &+\frac{\sin \theta \left(2 \gamma  M v-\alpha ^2 (\gamma -1) r \cos \theta\right)}{\alpha ^2 r^2} dr'\notag\\
 &+\left(\gamma  \sin ^2\theta+\cos ^2\theta\right) d\theta',\\
d\phi &= d\phi'.
\end{align}
The equivalence of \eqref{eq:edotinf} and \eqref{eq:pjetinf} for $v\ll 1$  follows from \eqref{eq:dtblboost}.  To see this, write $dE'=-\int T^{r'}_{t'}\sqrt{-g'}d\theta'dt'd\phi'=-\int T^{r}_{t'}\sqrt{-g} dt d\theta d\phi'$ and then note $dt=dt'+O(v)$.

The Schwarzschild components of the KS basis vectors are
\begin{align}
\partial_\tau &= \partial_t,\label{eq:dtauapp}\\
\partial_x &= -\frac{2M\sin\theta}{\alpha^2 r}\partial_t+\sin\theta \partial_r + \frac{\cos\theta}{r}\partial_\theta,\\
\partial_y &= \frac{1}{r\sin\theta}\partial_\theta,\\
\partial_z &= -\frac{2M\cos\theta}{\alpha^2 r}\partial_t+\cos\theta \partial_r - \frac{\sin\theta}{r}\partial_\theta.\label{eq:dzapp}
\end{align}
Equations \eqref{eq:utauapp}-\eqref{eq:uzapp} follow from these relations.

\section{Jet power at the horizon}
\label{sec:EMH}

Recall that the jet power \eqref{eq:edotinf} is
\beq\label{eq:edothor}
P_{\rm jet}  \equiv \frac{dE'}{dt'} = -\int_{S^2} T^{r'}_{t'} \sqrt{-g'} d\theta' d\phi'.
\eeq
The vector $\partial_{t'}$ is not Killing, so $P_{\rm jet}$ may be a function of radius.  In this section we evaluate the jet power at the horizon.  The astrophysically more interesting observable is the jet power at infinity, which we computed in Sec. \ref{sec:pjetinf}.  

As before, the FIDO-frame components of $T^r_{t'}$  are 
\beq\label{eq:trthor}
T^r_{t'} = \gamma \left(\alpha^2+\frac{2M}{r}v\cos\theta\right)T^{\hat{r}}_{\hat{t}}
-\gamma v \cos\theta T^{\hat{r}}_{\hat{r}}
+\gamma v \alpha \sin \theta T^{\hat{r}}_{\hat{\theta}},
\eeq
where $\alpha^2=1-2M/r$.
In the FIDO frame, the stress-energy tensor has its usual form \eqref{eq:t00}-\eqref{eq:tjk}.

At the horizon, the six components of the electromagnetic field are not all independent because radiation is always ingoing at the horizon.  In particular, we have the horizon boundary condition \cite{Parikh:1997ma}
\beq\label{eq:bchor}
\vec{E}_\parallel  = n \times \vec{B}_\parallel,
\eeq
where $\vec{E}_\parallel = (E^{\hat{\theta}},E^{\hat{\phi}})$, $\vec{B}_\parallel = (B^{\hat{\theta}},B^{\hat{\phi}})$, and $n$ is the outward-pointing unit normal vector.
In components, 
\beq
E^{\hat{\theta}}=-B^{\hat{\phi}},\quad E^{\hat{\phi}}=B^{\hat{\theta}}.
\eeq
This eliminates two components of the fields at the horizon.  We also have the force-free constraint $\mathbf{E}\cdot\mathbf{B}=0$.  Combined with the horizon boundary condition, it implies $E^{\hat{r}}=0$
or $B^{\hat{r}}=0$ at the horizon.  We choose $E^{\hat{r}}=0$. 

As before, we replace $\vec{E}_\parallel$ and $\vec{B}_\parallel$ with the field line velocity $\mathbf{v}_F$, defined by 
$\mathbf{E} = -\mathbf{v}_F\times \mathbf{B}$.  
In components, the fields at the horizon are
\begin{align}
E^{\hat{\phi}} &= B^{\hat{\theta}} = \frac{v^{\hat{\theta}}_F}{1+v^{\hat{r}}_F}B^{\hat{r}},\label{eq:ephor}\\
E^{\hat{\theta}} &= -B^{\hat{\phi}} = -\frac{v_F^{\hat{\phi}}}{1+v_F^{\hat{r}}}B^{\hat{r}}.
\end{align}
Plugging into \eqref{eq:trthor} gives the stress-energy tensor at the horizon,
\beq\label{eq:trthor2}
T^r_{t'}= \gamma \left(\frac{\alpha v_F^\parallel}{1+\vfr}\right)^2 B_{\hat{r}}^2
  +\frac{1}{2} \gamma v \cos\theta B_{\hat{r}}^2
  -\gamma v \sin\theta \frac{\alpha v^{\hat{\theta}}_F}{1+v^{\hat{r}}_F} B_{\hat{r}}^2,
\eeq
where $v_F^\parallel=\sqrt{(\vfh)^2+(\vfp)^2}$.  For small velocities,
\beq\label{eq:trthor2smallv}
T^r_{t'}\approx (\alpha v_F^\parallel)^2 B_{\hat{r}}^2
  +\frac{1}{2} \gamma v \cos\theta B_{\hat{r}}^2
  -\alpha v^{\hat{\theta}}_F v \sin\theta  B_{\hat{r}}^2.
\eeq
This is the same as the expression at infinity \eqref{eq:trtsimple}, except the first and third terms on the rhs differ by relative minus signs and by extra factors of $\alpha$.  At infinity, only the first term on the rhs contributed to the jet power, but at the horizon this term has the wrong sign to describe energy extraction.  It describes dissipation on the stretched horizon.  At the horizon, energy extraction is provided by the third term on the rhs of \eqref{eq:trthor2smallv}.  If we make the same assumptions as earlier for $B^{\hat{r}}$ and $\alpha \mathbf{v}_F$, we find $P_{\rm jet }= (v/2M)^2\Phi_H^2/(12\pi)$, where $\Phi_H = 2\pi r B_{\hat{r}}$ is the magnetic flux at the horizon.   As noted earlier, this need not match the jet power at infinity computed in Sec. \ref{sec:pjetinf} because $\partial_{t'}$ is not Killing.

\section{Outgoing boundary condition}
\label{sec:outgoing}

In Sec. \ref{sec:pjetinf}, we imposed the outgoing boundary condition \eqref{eq:bcinf}
\beq\label{eq:bcinf2}
\vec{E}_\parallel  = -n \times \vec{B}_\parallel
\eeq
at \ip.  This boundary condition has appeared before (see, e.g., \cite{2014ApJ...788..186N}).  It differs from the ingoing boundary condition imposed at black hole horizons by an overall minus sign.  A simple derivation of the horizon boundary condition has been given by \cite{1986bhmp.book.....T,Parikh:1997ma}.  In this section we adapt their argument to \ip\ and derive \eqref{eq:bcinf2}.  

Equation \eqref{eq:bcinf2} is expressed in the FIDO frame.  The FIDO frame is singular at \ip: all of \ip\ is mapped to $t=r=\infty$.  Outgoing Eddington-Finkelstein coordinates,
\beq
ds^2 = -\left(1-\frac{2M}{r}\right)du^2-2du dr+r^2 d\Omega^2,
\eeq 
are nonsingular there.  Lines of constant $u$ are null, but we can 
perturb them slightly so that they become timelike near \ip. Let $\tilde{\mathbf{E}}$ and $\tilde{\mathbf{B}}$ be the electric and magnetic fields measured by local observers in this frame.
$\tilde{\mathbf{E}}$ and $\tilde{\mathbf{B}}$ and the FIDO-frame fields are related by a Lorentz boost.  At stretched infinity, FIDOs move with velocity $v^{\hat{r}}\approx -1$ with respect to perturbed Eddington-Finkelstein observers, so
\begin{align}
E_{\hat{\theta}} &\approx \gamma(E_{\tilde{\theta}}+B_{\tilde{\phi}}),\label{eq:lorentz1}\\
E_{\hat{\phi}} &\approx \gamma(E_{\tilde{\phi}}-B_{\tilde{\theta}}),\\
B_{\hat{\theta}} &\approx \gamma(B_{\tilde{\theta}}-E_{\tilde{\phi}}),\\
B_{\hat{\phi}} &\approx \gamma(B_{\tilde{\phi}}+E_{\tilde{\theta}}).\label{eq:lorentz4}
\end{align}
If  $\tilde{\mathbf{E}}$ and $\tilde{\mathbf{B}}$ are finite, then it follows from \eqref{eq:lorentz1}-\eqref{eq:lorentz4} that $E_{\hat{\theta}}\approx B_{\hat{\phi}}$ and $E_{\hat{\phi}}\approx -B_{\hat{\theta}}$ on stretched infinity, with equality in the true infinity limit.  This proves \eqref{eq:bcinf2}.
The derivation of the ingoing boundary condition at the horizon is similar, except freely falling observers play the role of the perturbed Eddington-Finkelstein observers \cite{1986bhmp.book.....T,Parikh:1997ma}.

\bibliography{ms}

\end{document}